\documentclass[conference,a4paper]{IEEEtran}

\usepackage[utf8]{inputenc} 
\usepackage[T1]{fontenc}
\usepackage{url}
\usepackage{ifthen}
\usepackage{cite}
\usepackage[cmex10]{amsmath}
\usepackage{amsfonts}
\usepackage{amssymb}

\newtheorem{theorem}{Theorem}

\newtheorem{definition}[theorem]{Definition}

\newtheorem{proposition}[theorem]{Proposition}

\newenvironment{proof}[1][Proof]{\noindent\textbf{#1.} }{\ \rule{0.5em}{0.5em}}

\begin{document}
\title{Optimized quantum $f$-divergences} 


\author{
\IEEEauthorblockN{Mark M.~Wilde}
\IEEEauthorblockA{Hearne Institute for Theoretical Physics,
                    Department of Physics and Astronomy,
                    Center for Computation and Technology,\\
                    Louisiana State University,
                    Baton Rouge, Louisiana 70803, USA,
                    Email: mwilde@lsu.edu}
}


\maketitle

\begin{abstract}
The quantum relative entropy is a measure of the distinguishability of two
quantum states, and it is a unifying concept in quantum information theory:
many information measures such as entropy, conditional entropy, mutual
information, and entanglement measures can be realized from it. As such, there
has been broad interest in generalizing the notion to further understand its
most basic properties, one of which is the data processing inequality. The
quantum $f$-divergence of Petz is one generalization of the quantum relative
entropy, and it also leads to other relative entropies, such as the
Petz--R\'enyi relative entropies. In this contribution, I introduce the optimized
quantum $f$-divergence as a related generalization of quantum relative
entropy. I prove that it satisfies the data processing inequality, and the
method of proof relies upon the operator Jensen inequality, similar to Petz's
original approach. Interestingly, the sandwiched R\'enyi relative entropies
are particular examples of the optimized $f$-divergence. Thus, one benefit of
this approach is that there is now a single, unified approach for establishing
the data processing inequality for both the Petz--R\'enyi and sandwiched
R\'enyi relative entropies, for the full range of parameters for which it is
known to hold.

\end{abstract}

\textit{Full version of this paper is accessible at
arXiv:1710.10252
}

\section{Introduction}

The quantum relative entropy \cite{U62}\ is a foundational distinguishability
measure in quantum information theory (QIT). It is a function of two quantum states
and measures how well one can tell the two states apart by a quantum-mechanical experiment.
One important reason for why it has found such widespread application is that
it satisfies a data-processing inequality \cite{Lindblad1975,U77}: it does not
increase under the action of a quantum channel on the two states. This can be
interpreted as saying that two quantum states do not become more
distinguishable if the same quantum channel is applied to them, and a precise
interpretation of this statement in terms of quantum hypothesis testing is
available in \cite{HP91,ON00,BS12}. 
Quantum relative
entropy generalizes its classical counterpart \cite{kullback1951}.

The wide interest in relative entropy sparked various researchers to
generalize and study it further, in an attempt to elucidate the fundamental
properties that govern its behavior. One notable generalization is R\'enyi's
relative entropy \cite{Renyi61}, but this was subsequently generalized even
further in the form of the $f$-divergence \cite{C67,Ali1966,M63}. For
probability distributions $\{p(x)\}_{x}$ and $\{q(x)\}_{x}$ and a convex
function $f$, the $f$-divergence is defined as
$
\sum_{x}q(x)f(p(x)/q(x)),
$
in the case that $p(x)=0$ for all $x$ such that $q(x)=0$. The resulting
quantity is then non-increasing under the action of a classical channel
$r(y|x)$ that produces the output
distributions $\sum_{x}r(y|x)p(x)$ and $\sum_{x}r(y|x)q(x)$. Some years after
these developments, a quantum generalization of $f$-divergence appeared in
\cite{P85,P86}
In
\cite{P85,P86} and a later development \cite{TCR09}, the quantum
data-processing inequality was proved in full generality for arbitrary quantum
channels, whenever the underlying function $f$ is \textit{operator} convex. 

Interestingly, when generalizing a notion from classical to QIT, there is often more than one way to do so, and sometimes
there could even be an infinite number of ways to do so. This has to do with
the non-commutativity of quantum states.
For example, there are
several different ways that one could generalize the relative entropy to the
quantum case, and two prominent formulas were put forward in \cite{U62}\ and
\cite{Belavkin1982}. This added complexity for the quantum case could
potentially be problematic, but the typical way of determining on which
generalizations we should focus is to show that a given formula is the answer
to a meaningful operational task. The papers \cite{HP91,ON00} accomplished this for the formula from
\cite{U62}, and since then, researchers have
realized more and more just how foundational the formula of \cite{U62} is. As
a consequence, the formula of \cite{U62} is now known as quantum relative entropy.

The situation becomes more intricate when it comes to quantum generalizations
of R\'enyi relative entropy. For many years, the Petz--R\'enyi relative
entropy of \cite{P85,P86} has been widely studied and given an operational
interpretation \cite{N06,Hay07}, again in the context of quantum hypothesis
testing. 
However, in recent years,
the sandwiched R\'enyi relative entropy of \cite{MDSFT13,WWY13}\ has gained
prominence, due to its role in establishing strong converses for communication
tasks (see, e.g.,\cite{WWY13,WTB16}). The result of
\cite{MO13}\ solidified its fundamental meaning in QIT, proving that it has an operational interpretation in the
strong converse exponent of quantum hypothesis testing. As such, the situation
is that there are two generalizations of R\'enyi relative
entropy that should be considered in QIT, due to their
operational role mentioned above.

The same work that introduced the Petz--R\'enyi relative entropy also
introduced a quantum generalization of the notion of $f$-divergence
\cite{P85,P86} (see also \cite{HMPB11}), with the Petz--R\'enyi relative
entropy being a particular example. Since then, other quantum $f$-divergences
have appeared \cite{PR98,HM17}, now known as minimal and maximal
$f$-divergences \cite{M13,HM17}. However, it has not been known how
the sandwiched R\'enyi relative entropy fits into the paradigm of quantum
$f$-divergences.

In this paper, I modify Petz's definition of quantum $f$-divergence
\cite{P85,P86,HMPB11}, by allowing for a particular optimization (see
Definition~\ref{def:opt-f-div}\ for details of the modification). As such, I
call the resulting quantity the \textit{optimized} quantum $f$-divergence. I
prove that it obeys a quantum data processing inequality, and as such, my
perspective is that it deserves to be considered as another variant of the
quantum $f$-divergence, in addition to the original, the minimal, and the
maximal. Interestingly, the sandwiched R\'enyi relative entropy is directly
related to the optimized quantum $f$-divergence, thus bringing the sandwiched quantity into the
$f$-divergence formalism.

One benefit of the results of this paper is that there is now a single,
unified approach for establishing the data-processing inequality for both the
Petz--R\'enyi relative entropy and the sandwiched R\'enyi relative entropy,
for the full R\'enyi parameter ranges for which it is known to hold. This
unified approach is based on Petz's original approach that employed the
operator Jensen inequality \cite{HP03}, and it is useful for presenting a succint proof of
the data processing inequality for both quantum R\'enyi relative entropy families.

In the rest of the paper, I begin by defining the optimized quantum
$f$-divergence in the next section.
In Section~\ref{sec:DP}, I prove that the optimized $f$-divergence
satisfies the quantum data processing inequality under partial trace whenever the underlying
function $f$ is operator anti-monotone with domain $(0,\infty)$ and range
$\mathbb{R}$.
The core tool
underlying this proof is the operator Jensen inequality \cite{HP03}.
In Section~\ref{sec:examples-opt-f-div}, I show how
the quantum relative entropy and the sandwiched R\'enyi relative entropies are
directly related to the optimized quantum $f$-divergence.
Section~\ref{sec:petz-f}\ then discusses the relation between Petz's
$f$-divergence and the optimized one. 
I finally conclude in
Section~\ref{sec:conclusion}\ with a summary.

\section{Optimized quantum $f$-divergence}

Let us begin by defining the optimized quantum $f$-divergence. Here I focus exclusively on the case of positive definite operators, and the full version provides details for the more general case of positive semi-definite operators.

\begin{definition}
[Optimized quantum $f$-divergence]\label{def:opt-f-div}Let $f$ be a function
with domain $(0,\infty)$ and range $\mathbb{R}$. For positive definite
operators $X$ and $Y$ acting on a Hilbert space $\mathcal{H}_{S}$, we define
the \textit{optimized} quantum $f$-divergence as%
\begin{equation}
\widetilde{Q}_{f}(X\Vert Y)\equiv\sup_{\tau>0,\ \operatorname{Tr}\{\tau
\}\leq1}\widetilde{Q}_{f}(X\Vert Y;\tau), \label{eq:q-f-Y-not-invertible}%
\end{equation}
where $\widetilde{Q}_{f}(X\Vert Y;\tau)$ is defined for positive definite $Y$
and $\tau$ acting on $\mathcal{H}_{S}$ as%
\begin{align}
\widetilde{Q}_{f}(X\Vert Y;\tau)  &  \equiv\langle\varphi^{X}|_{S\hat{S}%
}f(\tau_{S}^{-1}\otimes Y_{\hat{S}}^{T})|\varphi^{X}\rangle_{S\hat{S}%
},\label{eq:q_f_tau}\\
|\varphi^{X}\rangle_{S\hat{S}}  &  \equiv(X_{S}^{1/2}\otimes I_{\hat{S}%
})|\Gamma\rangle_{S\hat{S}}.
\end{align}
In the above,
$\mathcal{H}_{\hat{S}}$ is an auxiliary Hilbert space isomorphic to
$\mathcal{H}_{S}$,
$
\left\vert \Gamma\right\rangle _{S\hat{S}}\equiv\sum_{i=1}^{\left\vert
S\right\vert }\left\vert i\right\rangle _{S}\left\vert i\right\rangle
_{\hat{S}},
$
for orthonormal bases $\{\left\vert i\right\rangle _{S}\}_{i=1}^{\left\vert
S\right\vert }$ and $\{\left\vert i\right\rangle _{\hat{S}}\}_{i=1}%
^{|\hat{S}|}$, and the $T$ superscript indicates transpose with respect to
the basis $\{\left\vert i\right\rangle _{\hat{S}}\}_{i}$.
\end{definition}

The case of greatest interest for us here is when the underlying function $f$
is operator anti-monotone; i.e., for Hermitian operators $A$ and $B$, the
function $f$ is such that $A\leq B\Rightarrow f(B)\leq f(A)$ (see, e.g.,
\cite{B97}). This property is rather strong, but there are several functions
of interest in quantum physical applications that obey it (see
Section~\ref{sec:examples-opt-f-div}). One critical property of an operator
anti-monotone function with domain $(0,\infty)$ and range $\mathbb{R}$ is that
it is also operator convex and continuous (see, e.g., \cite{H13}).
In this
case, we have the following simple proposition, proved in the full version:

\begin{proposition}
Let $f$ be an operator anti-monotone function with domain $(0,\infty)$ and
range $\mathbb{R}$. For positive definite operators $X$ and $Y$ acting on
a Hilbert space $\mathcal{H}_{S}$,
\begin{equation}
\widetilde{Q}_{f}(X\Vert Y)=\sup_{\tau>0,\, \operatorname{Tr}\{\tau\}=1}%
\widetilde{Q}_{f}(X\Vert Y;\tau),
\end{equation}
and  the function $\widetilde{Q}_{f}(X\Vert Y;\tau)$ is concave in $\tau$.
\end{proposition}

\section{Quantum data processing}

\label{sec:DP}
Our first main objective is to prove that $\widetilde{Q}%
_{f}(X\Vert Y)$ deserves the name \textquotedblleft$f$%
-divergence\textquotedblright\ or \textquotedblleft$f$-relative
entropy,\textquotedblright\ i.e., that it is monotone non-increasing under the
action of a completely positive trace-preserving map $\mathcal{N}$:%
\begin{equation}
\widetilde{Q}_{f}(X\Vert Y)\geq\widetilde{Q}_{f}(\mathcal{N}(X)\Vert
\mathcal{N}(Y)). \label{eq:mono-DP}%
\end{equation}
Such a map $\mathcal{N}$ is also called a quantum channel, due to its purpose
in quantum physics as modeling the physical evolution of the state of a
quantum system. In QIT contexts, the inequality in
\eqref{eq:mono-DP} is known as the quantum data processing inequality.
According to the Stinespring dilation theorem \cite{S55},
to
every quantum channel $\mathcal{N}_{S\rightarrow B}$, there exists an isometry
$U_{S\rightarrow BE}^{\mathcal{N}}$ such that%
\begin{equation}
\mathcal{N}_{S\rightarrow B}(X_{S})=\operatorname{Tr}_{E}\{U_{S\rightarrow
BE}^{\mathcal{N}}X_{S}\left(  U_{S\rightarrow BE}^{\mathcal{N}}\right)
^{\dag}\}.
\end{equation}
As such, we can prove the inequality in \eqref{eq:mono-DP} in two steps.
\textit{Isometric invariance}:\ First show that%
\begin{equation}
\widetilde{Q}_{f}(X\Vert Y)=\widetilde{Q}_{f}(UXU^{\dag}\Vert UYU^{\dag})
\end{equation}
for any isometry $U$ and any positive semi-definite $X$ and $Y$. This is done in the full version of this work, using the general definition given there.
 \textit{Monotonicity under partial trace}:\ Then show that%
\begin{equation}
\widetilde{Q}_{f}(X_{AB}\Vert Y_{AB})\geq\widetilde{Q}_{f}(X_{A}\Vert Y_{A})
\end{equation}
for positive semi-definite operators $X_{AB}$ and $Y_{AB}$ acting on the
tensor-product Hilbert space $\mathcal{H}_{A}\otimes\mathcal{H}_{B}$, with
$X_{A}=\operatorname{Tr}_{B}\{X_{AB}\}$ and $Y_{A}=\operatorname{Tr}%
_{B}\{Y_{AB}\}$.

We now discuss the second step toward quantum data processing, mentioned above, and here we focus exclusively on positive definite operators:

\begin{theorem}
[Monotonicity under partial trace]\label{thm:DP}
Let $f$ be an operator anti-monotone function with domain $(0,\infty)$ and
range $\mathbb{R}$. Given positive
definite operators $X_{AB}$ and $Y_{AB}$ acting on the tensor-product
Hilbert space $\mathcal{H}_{A}\otimes\mathcal{H}_{B}$, the optimized quantum
$f$-divergence does not increase under the action of a partial trace, in the
sense that%
\begin{equation}
\widetilde{Q}_{f}(X_{AB}\Vert Y_{AB})\geq\widetilde{Q}_{f}(X_{A}\Vert Y_{A}),
\label{eq:DP-partial-trace}%
\end{equation}
where $X_{A}=\operatorname{Tr}_{B}\{X_{AB}\}$ and $Y_{A}=\operatorname{Tr}%
_{B}\{Y_{AB}\}$.
\end{theorem}

\vspace{.1in}
\begin{proof}
%
The quantities
of interest are as follows:%
\begin{multline}
\widetilde{Q}_{f}(X_{AB}\Vert Y_{AB};\tau_{AB})    =\\
\langle\varphi^{X_{AB}%
}|_{AB\hat{A}\hat{B}}f(\tau_{AB}^{-1}\otimes Y_{\hat{A}\hat{B}}^{T}%
)|\varphi^{X_{AB}}\rangle_{AB\hat{A}\hat{B}},
\end{multline}
\begin{equation}
\widetilde{Q}_{f}(X_{A}\Vert Y_{A};\omega_{A})    =
\langle\varphi^{X_{A}%
}|_{A\hat{A}}f(\omega_{A}^{-1}\otimes Y_{\hat{A}}^{T})|\varphi^{X_{A}}%
\rangle_{A\hat{A}},
\end{equation}
where $\tau_{AB}$ and $\omega_{A}$ are invertible density operators and, by
definition,%
\begin{equation}
|\varphi^{X_{AB}}\rangle_{AB\hat{A}\hat{B}}=\left(  X_{AB}^{1/2}\otimes
I_{\hat{A}\hat{B}}\right)  |\Gamma\rangle_{A\hat{A}}\otimes|\Gamma
\rangle_{B\hat{B}}.
\end{equation}
The following map, acting on an operator $Z_{A}$, is a quantum channel known
as the Petz recovery channel \cite{Petz1986,Petz1988}:
\begin{equation}
Z_{A}\rightarrow X_{AB}^{1/2}\left(  \left[  X_{A}^{-1/2}Z_{A}X_{A}%
^{-1/2}\right]  \otimes I_{B}\right)  X_{AB}^{1/2}. \label{eq:petz-recovery}%
\end{equation}
It is completely positive because it consists of the serial concatenation of
three completely positive maps:\ sandwiching by $X_{A}^{-1/2}$, tensoring in
the identity $I_{B}$, and sandwiching by $X_{AB}^{1/2}$. It is also trace
preserving.
The Petz recovery channel has the property that it perfectly recovers $X_{AB}$
if $X_{A}$ is input because%
\begin{equation}
\!\!\!X_{A}\rightarrow X_{AB}^{1/2}\left(  \left[  X_{A}^{-1/2}X_{A}X_{A}%
^{-1/2}\right]  \otimes I_{B}\right)  X_{AB}^{1/2}=X_{AB}.
\label{eq:perfect-recovery}%
\end{equation}
Every completely positive and trace preserving map $\mathcal{N}$ has a Kraus
decomposition, which is a set $\{K_{i}\}_{i}$ of operators such that
$
\mathcal{N}(\cdot)=\sum_{i}K_{i}(\cdot)K_{i}^{\dag}$ and $\sum_{i}K_{i}^{\dag
}K_{i}=I.
$
A standard construction for an isometric extension of a channel is then to
pick an orthonormal basis $\{|i\rangle_{E}\}_{i}$ for an auxiliary Hilbert
space $\mathcal{H}_{E}$\ and define%
\begin{equation}
V=\sum_{i}K_{i}\otimes|i\rangle_{E}. \label{eq:iso-from-kraus}%
\end{equation}
One can then readily check that $\mathcal{N}(\cdot)=\operatorname{Tr}%
_{E}\{V(\cdot)V^{\dag}\}$ and $V^{\dag}V=I$. 
For the Petz recovery channel, we can
figure out a Kraus decomposition by expanding the identity operator
$I_{B}=\sum_{j=1}^{\left\vert B\right\vert }|j\rangle\langle j|_{B}$, with
respect to some orthonormal basis $\{|j\rangle_{B}\}_{j}$, so that%
\begin{align}
& X_{AB}^{1/2}\left(  \left[  X_{A}^{-1/2}\omega_{A}X_{A}^{-1/2}\right]  \otimes
I_{B}\right)  X_{AB}^{1/2}  \nonumber \\
&  =\sum_{j=1}^{\left\vert B\right\vert }%
X_{AB}^{1/2}\left(  \left[  X_{A}^{-1/2}\omega_{A}X_{A}^{-1/2}\right]
\otimes|j\rangle\langle j|_{B}\right)  X_{AB}^{1/2}\nonumber \\
&  =\sum_{j=1}^{\left\vert B\right\vert }X_{AB}^{1/2}\left[  X_{A}%
^{-1/2}\otimes|j\rangle_{B}\right]  \omega_{A}\left[  X_{A}^{-1/2}%
\otimes\langle j|_{B}\right]  X_{AB}^{1/2}.\nonumber 
\end{align}
Thus, Kraus operators for the Petz recovery channel are 
$
\left\{  X_{AB}^{1/2}\left[  X_{A}^{-1/2}\otimes|j\rangle_{B}\right]
\right\}  _{j=1}^{\left\vert B\right\vert }$.
According to the standard recipe in \eqref{eq:iso-from-kraus}, we can
construct an isometric extension of the Petz recovery channel\ as%
\begin{align}
\sum_{j=1}^{\left\vert B\right\vert }X_{AB}^{1/2}\left[  X_{A}^{-1/2}%
\otimes|j\rangle_{B}\right]  |j\rangle_{\hat{B}} 
 &  =X_{AB}^{1/2}X_{A}%
^{-1/2}\sum_{j=1}^{\left\vert B\right\vert }|j\rangle_{B}|j\rangle_{\hat{B}} \nonumber \\
&  =X_{AB}^{1/2}X_{A}^{-1/2}|\Gamma\rangle_{B\hat{B}}.
\end{align}
We can then extend this isometry to act as an isometry on a larger space by
tensoring it with the identity operator $I_{\hat{A}}$, and so we define%
\begin{equation}
V_{A\hat{A}\rightarrow A\hat{A}B\hat{B}}\equiv X_{AB}^{1/2}\left[
X_{A}^{-1/2}\otimes I_{\hat{A}}\right]  |\Gamma\rangle_{B\hat{B}}.
\end{equation}
We can also see that $V_{A\hat{A}\rightarrow A\hat{A}B\hat{B}}$ acting on
$|\varphi^{X_{A}}\rangle_{A\hat{A}}$ generates $|\varphi^{X_{AB}}%
\rangle_{AB\hat{A}\hat{B}}$:
$
|\varphi^{X_{AB}}\rangle_{AB\hat{A}\hat{B}}=V_{A\hat{A}\rightarrow A\hat
{A}B\hat{B}}|\varphi^{X_{A}}\rangle_{A\hat{A}}$.
This can be interpreted as a generalization of \eqref{eq:perfect-recovery} in
the language of QIT:\ an isometric extension of the Petz
recovery channel perfectly recovers a purification $|\varphi^{X_{AB}}%
\rangle_{AB\hat{A}\hat{B}}$ of $X_{AB}$ from a purification $|\varphi^{X_{A}%
}\rangle_{A\hat{A}}$ of $X_{A}$. Since the Petz recovery channel is indeed a
channel, we can pick $\tau_{AB}$ as the output state of the Petz recovery
channel acting on an invertible state $\omega_{A}$:%
\begin{equation}
\tau_{AB}=X_{AB}^{1/2}\left(  \left[  X_{A}^{-1/2}\omega_{A}X_{A}%
^{-1/2}\right]  \otimes I_{B}\right)  X_{AB}^{1/2}.
\label{eq:tau-state-choice}%
\end{equation}
Observe that $\tau_{AB}$ is invertible. Then consider that%
\begin{align}
&  V^{\dag}\left(  \tau_{AB}^{-1}\otimes Y_{\hat{A}\hat{B}}^{T}\right)
V\nonumber\\
&  =\langle\Gamma|_{B\hat{B}}\Big(  X_{A}^{-1/2}X_{AB}^{1/2}\tau_{AB}%
^{-1}X_{AB}^{1/2}X_{A}^{-1/2}\otimes Y_{\hat{A}\hat{B}}^{T}\Big)
|\Gamma\rangle_{B\hat{B}}\\
&  =\langle\Gamma|_{B\hat{B}}\left(  \omega_{A}^{-1}\otimes I_B \otimes Y_{\hat{A}\hat{B}%
}^{T}\right)  |\Gamma\rangle_{B\hat{B}}\\
&  =\omega_{A}^{-1}\otimes\langle\Gamma|_{B\hat{B}}Y_{\hat{A}\hat{B}}%
^{T}|\Gamma\rangle_{B\hat{B}}\\
&  =\omega_{A}^{-1}\otimes Y_{\hat{A}}^{T}.
\end{align}
For the second equality, we used the fact that $X_{A}%
^{-1/2}X_{AB}^{1/2}\tau_{AB}^{-1}X_{AB}^{1/2}X_{A}^{-1/2}=\omega_{A}^{-1} \otimes I_{B}$ for the choice of $\tau_{AB}$ in \eqref{eq:tau-state-choice}.
With this setup, we can now readily establish the desired inequality by
employing the operator Jensen inequality \cite{HP03} and operator convexity of
the function~$f$:%
\begin{align}
& \widetilde{Q}_{f}(X_{AB}\Vert Y_{AB};\tau_{AB})  \nonumber \\
&  =\langle\varphi^{X_{AB}%
}|_{AB\hat{A}\hat{B}}f(\tau_{AB}^{-1}\otimes Y_{\hat{A}\hat{B}}^{T}%
)|\varphi^{X_{AB}}\rangle_{AB\hat{A}\hat{B}}\label{eq:op-convex-1}\\
&  =\langle\varphi^{X_{A}}|_{A\hat{A}}V^{\dag}f(\tau_{AB}^{-1}\otimes
Y_{\hat{A}\hat{B}}^{T})V|\varphi^{X_{A}}\rangle_{A\hat{A}}\\
&  \geq\langle\varphi^{X_{A}}|_{A\hat{A}}f(V^{\dag}[\tau_{AB}^{-1}\otimes
Y_{\hat{A}\hat{B}}^{T}]V)|\varphi^{X_{A}}\rangle_{A\hat{A}}\\
&  =\langle\varphi^{X_{A}}|_{A\hat{A}}f(\omega_{A}^{-1}\otimes Y_{\hat{A}}%
^{T})|\varphi^{X_{A}}\rangle_{A\hat{A}}\\
&  =\widetilde{Q}_{f}(X_{A}\Vert Y_{A};\omega_{A}). \label{eq:op-convex-last}%
\end{align}
Taking a supremum over $\tau_{AB}$ such that $\tau_{AB}>0$ and
$\operatorname{Tr}\{\tau_{AB}\}=1$, we conclude that the following inequality
holds for all invertible states $\omega_{A}$:%
\begin{equation}
\widetilde{Q}_{f}(X_{AB}\Vert Y_{AB})\geq\widetilde{Q}_{f}(X_{A}\Vert
Y_{A};\omega_{A}).
\end{equation}
After taking a supremum over invertible states $\omega_{A}$, we find that the
inequality in \eqref{eq:DP-partial-trace} holds when $X_{AB}$ is invertible.
\end{proof}

\section{Examples of optimized quantum $f$-divergences}

\label{sec:examples-opt-f-div}I now show how several known quantum divergences
are particular examples of an optimized quantum $f$-divergence, including the
quantum relative entropy \cite{U62}\ and the sandwiched R\'{e}nyi relative
quasi-entropies \cite{MDSFT13,WWY13}. The result will be that Theorem~\ref{thm:DP}
recovers quantum data processing for the sandwiched R\'{e}nyi relative
entropies for the full range of parameters for which it is known to hold.
Thus, one benefit of Theorem~\ref{thm:DP} and earlier work of
\cite{P85,P86,TCR09}\ is a single, unified approach, based on the operator
Jensen inequality \cite{HP03}, for establishing quantum data processing for
all of the Petz-- and sandwiched R\'{e}nyi relative entropies for the full
parameter ranges for which data processing is known to hold.

\subsection{Quantum relative entropy as optimized quantum $f$-divergence}

\label{sec:q-rel-ent}Let $\tau$ be an invertible state and
$X$ and $Y$ positive definite. Let $\overline{X}=X/\operatorname{Tr}%
\{X\}$. Pick the function
$
f(x)=-\log x $,
which is an operator anti-monotone function with domain $(0,\infty)$ and range
$\mathbb{R}$, and we find that%
\begin{align}
&  \frac{1}{\operatorname{Tr}\{X\}}\langle\varphi^{X}|_{S\hat{S}}\left[
-\log(\tau_{S}^{-1}\otimes  Y
_{\hat{S}}^{T})\right]  |\varphi^{X}\rangle_{S\hat{S}}\nonumber\\
&  =\langle\varphi^{\overline{X}}|_{S\hat{S}}\left[  \log(\tau_{S})\otimes
I_{\hat{S}}-I_{S}\otimes\log  Y
_{\hat{S}}^{T}\right]  |\varphi^{\overline{X}}\rangle_{S\hat{S}}\\
&  =\langle\varphi^{\overline{X}}|_{S\hat{S}}\log(\tau_{S})\otimes I_{\hat{S}%
}|\varphi^{\overline{X}}\rangle_{S\hat{S}}\nonumber \\
&\qquad -\langle\varphi^{\overline{X}%
}|_{S\hat{S}}I_{S}\otimes\log\left(  Y
_{\hat{S}}^{T}\right)|\varphi^{\overline{X}}\rangle_{S\hat{S}}\\
&  =\operatorname{Tr}\{\overline{X}\log\tau\}-\operatorname{Tr}\{\overline
{X}\log Y \}\\
&  \leq\operatorname{Tr}\{\overline{X}\log\overline{X}\}-\operatorname{Tr}%
\{\overline{X}\log Y \}
  =D(\overline{X}\Vert Y).
\end{align}
The inequality is a consequence of Klein's inequality \cite{O31} (see also
\cite{R02}), establishing that the optimal $\tau$ is set to $\overline{X}%
$. 
 So we find that
$
\widetilde{Q}_{-\log(\cdot)}(X\Vert Y)=\operatorname{Tr}\{X\}D(\overline
{X}\Vert Y),
$
where the quantum relative entropy $D(\overline{X}\Vert Y)$ is defined as
\cite{U62}
$
D(\overline{X}\Vert Y)=\operatorname{Tr}\{\overline{X}\left[  \log\overline
{X}-\log Y\right]  \}$.


\subsection{Sandwiched R\'enyi relative quasi-entropy as optimized quantum
$f$-divergence}

Take $\tau$, $X$, and $Y$ as defined in
Section~\ref{sec:q-rel-ent}. For $\alpha\in\lbrack1/2,1)$, pick the function
$
f(x)=-x^{\left(  1-\alpha\right)  /\alpha},
$
which is an operator anti-monotone function with domain $(0,\infty)$ and range
$\mathbb{R}$. Note that this is a reparametrization of $-x^{\beta}$ for
$\beta\in(0,1]$. I now show that%
\begin{equation}
\widetilde{Q}_{-\left(  \cdot\right)  ^{\left(  1-\alpha\right)  /\alpha}%
}(X\Vert Y)=-\left\Vert Y^{\left(  1-\alpha\right)  /2\alpha}XY^{\left(
1-\alpha\right)  /2\alpha}\right\Vert _{\alpha},
\end{equation}
which is the known expression for sandwiched quasi-entropy
for $\alpha\in\lbrack1/2,1)$ \cite{MDSFT13,WWY13}.
To see this, consider that%
\begin{align}
&  -\langle\varphi^{X}|_{S\hat{S}}\left[  \tau_{S}^{-1}\otimes
Y  _{\hat{S}}^{T}\right]  ^{\left(
1-\alpha\right)  /\alpha}|\varphi^{X}\rangle_{S\hat{S}}\nonumber\\
&  =-\langle\varphi^{X}|_{S\hat{S}}\tau_{S}^{\left(  \alpha-1\right)  /\alpha
}\otimes\left(    Y  _{\hat{S}}%
^{T}\right)  ^{\left(  1-\alpha\right)  /\alpha}|\varphi^{X}\rangle_{S\hat{S}%
}\nonumber\\
&  =-\langle\Gamma|_{S\hat{S}}X_{S}^{1/2}\tau_{S}^{\left(  \alpha-1\right)
/\alpha}X_{S}^{1/2}\otimes\left(    Y
_{\hat{S}}^{T}\right)  ^{\left(  1-\alpha\right)  /\alpha}|\Gamma
\rangle_{S\hat{S}}\nonumber\\
&  =-\operatorname{Tr}\left\{  X^{1/2}\tau^{\left(  \alpha-1\right)  /\alpha
}X^{1/2}Y^{\left(  1-\alpha\right)  /\alpha
}\right\}  \nonumber\\
&  =-\operatorname{Tr}\left\{  X^{1/2}Y^{\left(
1-\alpha\right)  /\alpha}X^{1/2}\tau^{\left(  \alpha-1\right)  /\alpha
}\right\}  .
\end{align}
Now optimizing over invertible states $\tau$ and employing H\"{o}lder duality,
in the form of the reverse H\"{o}lder inequality and as observed in
\cite{MDSFT13}, we find that
\begin{multline}
\sup_{\substack{\tau>0,\\\operatorname{Tr}\{\tau\}=1}}\left[
-\operatorname{Tr}\left\{  X^{1/2}Y^{\frac{
1-\alpha}{  \alpha}}X^{1/2}\tau^{\frac{  \alpha-1}{  \alpha}
}\right\}  \right]  \\
=-\left\Vert X^{1/2}Y^{\left(  1-\alpha\right)  /\alpha}X^{1/2}\right\Vert _{\alpha},
\end{multline}
where for positive semi-definite $Z$, we define
$
\left\Vert Z\right\Vert _{\alpha}=\left[  \operatorname{Tr}\{Z^{\alpha
}\}\right]  ^{1/\alpha}$.
We then get that%
\begin{align}
\widetilde{Q}_{-\left(  \cdot\right)  ^{\left(  1-\alpha\right)  /\alpha}%
}(X\Vert Y) & =-\left\Vert X^{1/2}Y^{\left(  1-\alpha\right)  /\alpha}%
X^{1/2}\right\Vert _{\alpha}\\
& =-\left\Vert Y^{\left(  1-\alpha\right)  /2\alpha
}XY^{\left(  1-\alpha\right)  /2\alpha}\right\Vert _{\alpha},
\end{align}
which is the sandwiched R\'{e}nyi relative quasi-entropy for the range
$\alpha\in\lbrack1/2,1)$. The sandwiched R\'{e}nyi relative entropy itself is
defined up to a normalization factor as \cite{MDSFT13,WWY13}
\begin{equation}
\widetilde{D}_{\alpha}(X\Vert Y)=\frac{\alpha}{\alpha-1}\log\left\Vert
Y^{\left(  1-\alpha\right)  /2\alpha}XY^{\left(  1-\alpha\right)  /2\alpha
}\right\Vert _{\alpha}.\label{eq:sandwiched-Renyi}%
\end{equation}
Thus, Theorem~\ref{thm:DP}\ implies quantum data processing for the sandwiched
R\'{e}nyi relative entropy
$
\widetilde{D}_{\alpha}(X_{AB}\Vert Y_{AB})\geq\widetilde{D}_{\alpha}(
X_A\Vert Y_A),
$
for the parameter range $\alpha\in\lbrack1/2,1)$, which is a result previously
established in \cite{FL13}.

For $\alpha\in(1,\infty]$, pick the function
$
f(x)=x^{\left(  1-\alpha\right)  /\alpha},
$
which is an operator anti-monotone function with domain $(0,\infty)$ and range
$\mathbb{R}$. Note that this is a reparametrization of $x^{\beta}$ for
$\beta\in\lbrack-1,0)$. I now show that%
\begin{equation}
\widetilde{Q}_{\left(  \cdot\right)  ^{\left(  1-\alpha\right)  /\alpha}%
}(X\Vert Y)=
\left\Vert Y^{\left(  1-\alpha\right)  /2\alpha}XY^{\left(  1-\alpha\right)
/2\alpha}\right\Vert _{\alpha}
,
\end{equation}
which is the known expression for sandwiched R\'{e}nyi relative quasi-entropy
for $\alpha\in(1,\infty]$ \cite{MDSFT13,WWY13}. To see this, consider that the same development as
above gives that%
\begin{multline}
\langle\varphi^{X}|_{S\hat{S}}(\tau_{S}^{-1}\otimes  Y  _{\hat{S}}^{T})^{\left(  1-\alpha\right)  /\alpha
}|\varphi^{X}\rangle_{S\hat{S}}
\\
=\operatorname{Tr}\left\{  X^{1/2}%
Y^{\left(  1-\alpha\right)  /\alpha}X^{1/2}%
\tau^{\left(  \alpha-1\right)  /\alpha}\right\}  .
\end{multline}
Again employing H\"{o}lder duality, as observed in \cite{MDSFT13}, we find
\begin{multline}
\sup_{\tau>0,\operatorname{Tr}\{\tau\}=1}\operatorname{Tr}\left\{
X^{1/2}Y^{\left(  1-\alpha\right)  /\alpha
}X^{1/2}\tau^{\left(  \alpha-1\right)  /\alpha}\right\}  \\
=\left\Vert
X^{1/2}Y^{\left(  1-\alpha\right)  /\alpha
}X^{1/2}\right\Vert _{\alpha},
\end{multline}
We then get that%
\begin{align}
\widetilde{Q}_{\left(  \cdot\right)  ^{\left(  1-\alpha\right)  /\alpha}%
}(X\Vert Y)& =\left\Vert X^{1/2}Y^{\left(  1-\alpha\right)  /\alpha}%
X^{1/2}\right\Vert _{\alpha}\\
& =\left\Vert Y^{\left(  1-\alpha\right)  /2\alpha
}XY^{\left(  1-\alpha\right)  /2\alpha}\right\Vert _{\alpha},
\end{align}
where the equalities hold
as observed in
\cite{MDSFT13}. The sandwiched R\'{e}nyi relative entropy itself is defined up
to a normalization factor as in \eqref{eq:sandwiched-Renyi}
\cite{MDSFT13,WWY13}. Thus, Theorem~\ref{thm:DP}\ implies quantum data
processing for the sandwiched R\'{e}nyi relative entropy
$
\widetilde{D}_{\alpha}(X_{AB}\Vert Y_{AB})\geq\widetilde{D}_{\alpha}(
X_A\Vert Y_A),
$
for the parameter range $\alpha\in(1,\infty]$, which is a result previously
established in full by \cite{FL13,B13monotone,MO13} and for $\alpha\in(1,2]$
by \cite{MDSFT13,WWY13}.

\section{On Petz's quantum $f$-divergence}

\label{sec:petz-f}
I now discuss in more detail the relation between the
optimized quantum $f$-divergence and Petz's $f$-divergence from
\cite{P85,P86}. In brief, we find that the Petz $f$-divergence can be
recovered by replacing $\tau$ in Definition~\ref{def:opt-f-div} with
$X$.

\begin{definition}
[Petz quantum $f$-divergence]\label{def:petz-f-div}Let $f$ be a continuous
function with domain $(0,\infty)$ and range $\mathbb{R}$. For positive
definite operators $X$ and $Y$ acting on a Hilbert space $\mathcal{H}%
_{S}$, the Petz quantum $f$-divergence is defined as%
\begin{equation}
Q_{f}(X\Vert Y)\equiv\langle
\varphi^{X}|_{S\hat{S}}f\left(    X_{S}
^{-1}\otimes  Y_{\hat{S}}
^{T}\right)  |\varphi^{X}\rangle_{S\hat{S}},
\end{equation}
where 
the notation
is the same as in Definition$~$\ref{def:opt-f-div}.
\end{definition}


One main concern is about quantum data processing with the Petz $f$-divergence. To show this, we take $f$ to be an operator anti-monotone
function with domain $(0,\infty)$ and range $\mathbb{R}$. As discussed in
Section~\ref{sec:DP}, one can establish data processing by showing isometric
invariance and monotonicity under partial trace. Isometric invariance of
$Q_{f}(X\Vert Y)$ follows from the same proof as given in
the full version and was also shown in \cite{TCR09}. Monotonicity of $Q_{f}(X_{AB}\Vert Y_{AB})$
under partial trace for positive definite
$X_{AB}$ and $Y_{AB}$ follows from the operator Jensen inequality \cite{P85,P86}.

Special and interesting cases of the Petz $f$-divergence are found by taking
$
f(x)    =-\log x,$ 
$f(x)    =-x^{\beta}$ for $\beta\in(0,1]$,
and $f(x)    =x^{\beta}$  for $\beta\in\lbrack-1,0)$.
Each of these functions are operator anti-monotone with domain $(0,\infty)$
and range $\mathbb{R}$.
As shown in \cite{P85,P86}, all of the following
quantities obey the data processing inequality:%
\begin{align}
Q_{-\log(\cdot)}(X\Vert Y)  &  =\operatorname{Tr}\{X\}D(\overline{X}\Vert
Y),\\
Q_{-(\cdot)^{\beta}}(X\Vert Y)  &  =-\operatorname{Tr}\{X^{1-\beta}Y^{\beta
}\},\text{ for }\beta\in(0,1],\\
Q_{(\cdot)^{\beta}}(X\Vert Y)  &  =
\operatorname{Tr}\{X^{1-\beta}Y^{\beta}\}   ,\ \text{for }\beta\in\lbrack-1,0).
\end{align}
By a reparametrization $\alpha=1-\beta$, the latter two
quantities are directly related to the Petz R\'enyi relative entropy
$
D_{\alpha}(X\Vert Y)\equiv
\frac{1}{\alpha-1}\log\operatorname{Tr}\{X^{\alpha}Y^{1-\alpha}\} 
$.
Thus, the data processing inequality holds for
$D_{\alpha}(X\Vert Y)$ for $\alpha\in\lbrack0,1)\cup(1,2]$  \cite{P86,TCR09}.

\section{Conclusion}

\label{sec:conclusion}The main contribution of the present work is the
definition of the optimized quantum $f$-divergence and the proof that the data
processing inequality holds for it whenever the function $f$ is operator
anti-monotone with domain $(0,\infty)$ and range $\mathbb{R}$. The proof of
the data processing inequality relies on the operator Jensen inequality
\cite{HP03}, and it bears some similarities to the original approach from
\cite{P85,P86,TCR09}. Furthermore, I showed how the sandwiched R\'enyi
relative entropies are particular examples of the optimized quantum
$f$-divergence. As such, one benefit of this paper is that there is now a
single, unified approach, based on the operator Jensen inequality \cite{HP03},
for establishing the data processing inequality for the Petz--R\'enyi and
sandwiched R\'enyi relative entropies, for the full range of parameters for
which it is known to hold.

%

\textbf{Acknowledgements.} I thank Anna Vershynina for discussions related to
the topic of this paper, and I acknowledge support from the NSF under grant no.~1714215.

\bibliographystyle{IEEEtran}
\bibliography{Ref}

\end{document}